\documentclass[twocolumn,showpacs,preprintnumbers,amsmath,amssymb,superscriptaddress,floatfix]{revtex4}
\usepackage[dvips]{graphicx}
\usepackage{dcolumn}
\usepackage{bm}
\usepackage{latexsym, amssymb,amsmath,amscd, psfrag}
\newcommand{\figwidth}{1\columnwidth}

\newcommand{\since}{.^{\kern-6.5pt{\textstyle\cdot\kern5pt\cdot}}~~}

\usepackage{eucal}%
\input amssym.def
\begin{document}
\title{Calculation of Atomic Number States: a Bethe Ansatz Approach}
\author{Shou-Pu Wan}
\email[email: ]{swan@physics.utexas.edu} \affiliation{Department
of physics, The University of Texas, Austin, TX 78712}
\affiliation{Center for Nonlinear Dynamics, The University of
Texas, Austin, TX 78712}
\author{Mark G. Raizen}
\affiliation{Department of physics, The University of Texas,
Austin, TX 78712} \affiliation{Center for Nonlinear Dynamics, The
University of Texas, Austin, TX 78712}
\author{Qian Niu}
\affiliation{Department of physics, The University of Texas,
Austin, TX 78712}
\date{\today}
\begin{abstract}
We analyze the conditions for producing atomic number states in a
one-dimensional optical box using the Bethe ansatz method. This
approach provides a general framework, enabling the study of
number state production over a wide range of realistic
experimental parameters.
\end{abstract}


\maketitle


\section{Introduction \label{sec:intro}}
The realization of Bose-Einstein condensation (BEC) in dilute
gases has enabled the study and control of many-body systems.
While most of the work has focused on the properties and
excitations of the condensate, it has provided a new path towards
generation of atomic number (or Fock) states. These few body
states with a definite number of atoms in the ground state are of
great interest for quantum information where individual qubits can
be addressed \cite{Dudarev:2003,Jaksch:1999,Andersson:2000}, and
could also be important for atom interferometry in order to reach
the Heisenberg limit of detection. Experimentally, realization of
Fock states requires a BEC confined in an optical box coupled with
single-atom counting. The challenge is to obtain confinement in a
trap that is comparable to optical lattices, but with only a
single site \cite {Meyrath:2005}. Recent experimental work has
demonstrated all the necessary steps towards this goal and they
are now being incorporated into a single system \cite{Chuu:2005}.
In parallel, the theoretical analysis of this problem has focused
on the conditions for optimum number state production
\cite{Dudarev:2007,ACampo:2008-2009}. These include the role of
varying trap depth and size, either separately or in tandem. The
interaction strength is a third control parameter that can be
tuned with the the help of Feshbach resonances or by tuning the
transverse confinement of the optical trap
\cite{Meyrath:2005,Inouye:1998} and is considered here in more
detail.

It is clear that strong repulsion between atoms are desirable for
the production of number states. The infinitely strong interaction
regime is the so-called Tonks-Girardeau regime where calculation
has been made trivial thanks to the boson-fermion correspondence
\cite{Girardeau:1960}. However, this is only an unreachable
limiting case. To make realistic predictions about what can be
experimentally realized, we must consider the regime of relatively
but not infinitely strong interactions. Previous studies of this
regime were carried out using numerical methods which are both
time-consuming and reveal little insight on how number states vary
with interaction strength. In this paper we develop an approach
that takes interaction as a parameter and by which we are able to
chart number states in the parameter space with interaction as one
of its dimensions \cite{comment1}.

Our analysis of the production of atomic number states in a 1D
optical box is based on the so-called Bethe ansatz approach. This
method was first developed by Hans Bethe to solve the problem of a
one-dimensional (1D) spin 1/2 Heisenberg ferromagnet
\cite{Bethe:1931}. Since it's invention, Bethe's method has found
important applications in the study of interacting spin systems
\cite{Orbach:1958,Walker:1959,Lieb:Schultz:Mattis:1961}. It has
also been applied to solve the problem of a 1D bosonic gas with
repulsive $\delta$-function interactions
\cite{Lieb:Liniger:1963,McGuire:1964,Yang:1967,Gaudin:1967,Sutherland:1968,Li:1995,Hao:2006}.
An outline for the structure of this article is as follows. In
section \ref{sec:formul}, we formulate the problem; in section
\ref{sec:betheansatz}, we present approximate solutions to our
problem with Bethe ansatz approach; in section
\ref{sec:number-state}, we use Bethe ansatz solutions to analyze
issues related to number state production.
\section{Formulation of the problem \label{sec:formul}}
The problem of many bosons with a $\delta$-function interaction
trapped in 1D square well potential with finite well depth was
studied using the Bethe ansatz in Ref \cite{Li:1995}. We find the
problem of producing Fock states in the ultracold atom systems
trapped in 1D optical box bears similar characteristics. We treat
the 1D optical trap as a square well potential of length $L$ and
depth $V_0$. We write the interaction potential as
$\frac{\hbar^2}{m}c \delta(x_i - x_j)$, where $x_i$ and $x_j$ are
the positions of the interacting particles, $\hbar$ is Planck's
constant and $m$ is an atom's mass, and $c$ is the interaction
strength and has dimension of $[1/\mbox{length}]$. According to
\cite{Olshanii:1998} we have the following expression for $c$,
\begin{equation}\label{inter:c}
c = \frac{4a}{{a_\perp}^2} \left(1-C \frac{a}{a_\bot}
\right)^{-1},
\end{equation}
where $a$ is the s-wave scattering length in 3-dimensional space,
$a_\perp = \sqrt{ {2\hbar^2}/ m\omega _\perp}$, $\omega_\perp$ is
the transverse trapping frequency, and $C \approx 1.4603$ is an
empirical constant number. Since the interaction strength $c$
depends on both scattering length and transverse trapping
frequency $\omega_{\perp}$, tuning either of them will affect the
interaction strength. The transverse trapping frequency may be
controlled by optical box parameters \cite {Meyrath:2005}.
Scattering length may be adjusted by Feshbach resonance \cite
{Inouye:1998}. To give a sense of order of magnitude, for sodium
atoms trapped in a 1D optical box with transverse trapping
frequency $\omega_\perp=2\pi \times 150$ kHz and zero magnetic
field, we have $c = 16863.6 \mbox{ cm}^{-1}$. For $^{87}$Rb atoms
in a similar trap with zero magnetic field, we have $c =92391.6
\mbox{ cm}^{-1}$.

To make our equations dimensionless, we use $1/c$ as the length
unit and $\hbar^2c^2/m$ as the energy unit. The square well
potential is then
\begin{equation} \label{square-well_pot} V\left(x\right) = \left\{
\begin{array}{cc}
  -{k_0}^2/2, & |x| < x_0/2 ,\\
  0, & \mbox{ otherwise} ,\\
\end{array}
\right.
\end{equation}
where $k_0$ and $x_0$ are dimensionless numbers. With these
parameters, the well width is $L = x_0/c$ and well depth is $V_0 =
\hbar^2c^2{k_0}^2/2m$ in cgs unit.

The hamiltonian for the many-body system may be written as
\begin{equation}\label{N-body:hamil}
{\mathcal H} = -\frac{1}{2} \sum_{i=1}^{N}{\frac{\partial^2}
{\partial x_i^2}} +\sum_{i=1}^{N}{V\left (x_i \right)} +
\mathop{\sum_{i,j=1}^{N}}_{i>j} {\delta \left( x_i - x_j \right)}.
\end{equation}
Our first main step is to solve the following eigenvalue problem
\begin{equation}\label{eigen:problem}
{\mathcal H} \psi(\overrightarrow x) = E \psi(\overrightarrow x),
\end{equation} where $\overrightarrow x$ is shorthand for
$x_1,x_2,\cdots, x_N$.
We are primarily interested in bound states whose wavefunctions
must be normalizable. As a minimum requirement, the wavefunction
of a bound state must satisfy $\lim_{x\to \pm \infty} {\psi
(\overrightarrow x)} = 0$.
\section{Bethe Ansatz solutions \label{sec:betheansatz}}
In section \ref{sec:formul}, we defined a boundary value problem
relevant to the production of number states. We will in this
section apply Bethe ansatz and obtain solutions for it.

As studied in previous literatures
\cite{Li:1995,Yang:1967,Lieb:Liniger:1963}, Bethe ansatz solution
of this problem introduces a set of $N$ as-yet-unknown wave
numbers $\overrightarrow {k}=\{k_1,k_2, \cdots, k_N\}$. In
conjugate to these wave numbers, another set $\overrightarrow
{\kappa}= \{\kappa_1,\kappa_2, \cdots, \kappa_N \}$ is defined as
\begin{equation} \label{define:kappa}
\kappa_{j}= \sqrt{k_0^2 - {k_{j}}^2},
\end{equation} for $j = 1,2,\cdots, N$. The total energy of the Bethe
ansatz state is $E= - \sum_{j}^{N}{ {\kappa_{j}}^2/2}
=\sum_{j}^{N}{ \left({k_j}^2-{k_0}^2 \right)/2} $. The
eigenfunction (wavefunction) of Eq.\eqref{eigen:problem} is
piecewise continuous in the $N$-dimensional coordinate space
$\{x_1,x_2,\cdots, x_N\}$. For simplicity, we consider three
representative regions in the $N$-dimensional coordinate space:
\begin{eqnarray}
R_1&:& -x_0/2 <x_1 < x_2 < \cdots < x_N < x_0/2,\\
R_2&:& x_1 < -x_0/2 < x_2 < \cdots < x_N < x_0/2, \\
R_3&:& -x_0/2 <x_1 < \cdots < x_{N-1} < x_0/2 < x_N.
\end{eqnarray}
$R_1$ represents a region where all particles are trapped; $R_{
2,( 3)}$ represent a region where the $1$st ($N$th) particle
tunnels into the left (right) barrier. In fact, each of these
regions falls in a class consisting of $N!$ regions that are
related by coordinate permutation. For ease of reference, we name
$\mathcal{A}$ the class of regions that can be obtained from $R_1$
by mere coordinate permutations and study the wavefunctions in
these regions at once.

We denote the wavefunctions in a region of $\mathcal{A}$ as
$\phi_\tau(\overrightarrow x)$, where $\tau$ is the permutation
operator that transforms $R_1$ into this region, \textit{i.e.},
$R_\tau = \tau R_1$. This wavefunction is the superposition of
pure plane waves with ``$\pm$ signs'' times permuted wave numbers,
\begin{equation}\label{Bethe:Ansatz:wavefunction:tau}
\phi_{ \mathbf{ \tau}}( \overrightarrow x) = \sum_{\varsigma \in
C_2^N}{\sum_{\mathbf{\sigma} \in \mathbf{G}}{A\left( \varsigma,
\sigma; \mathbf{\tau}\right) e^{ i( \varsigma \mathbf{ \sigma}
\overrightarrow{k}) \cdot \overrightarrow{x}}}}.
\end{equation}
where $\varsigma \equiv \left\{\varsigma_1, \varsigma_2, \cdots,
\varsigma_N \right\}$ represents a possible combination of $N$
signs each of which is either $+$ or $-$ and $C_2^N$ represents
the group of such operations, $\mathbf{G}$ is the permutation
group of $N$ particles, and $A(\varsigma, \mathbf{\sigma}; \tau)$
is the superposition amplitude.

It is clear that the superposition amplitude $A(\varsigma,
\mathbf{\sigma}; \mathbf{\tau})$ is a functional of the
sign-flipping operator, the wave number permutation operator, and
the region permutation operator. By bosonic particle permutation
symmetry, we establish the first set of equations among the
superposition amplitudes,
\begin{equation}\label{boson:exchange:symmetry}
A(\varsigma, \mathbf{\sigma}; \mathbf{\tau}) = A(\varsigma,
\mathbf{\tau\sigma}; \mathbf{I}),
\end{equation} where $\mathbf{I}$ is the identity element in the
permutation group.

The wavefunctions in region $R_2$ have a more complicated form,
\begin{equation}\label{Bethe:Ansatz:wavefunction:2}
\phi_{2}(\overrightarrow x)=
\sum_{\varsigma}{\sum_{\mathbf{\sigma} \in \mathbf{G}}
{B(\varsigma,\mathbf{\sigma}) e^{{(\mathbf{ \sigma}
\overrightarrow{\kappa})}_1 x_1} e ^{ i
\sum_{j=2}^N{(\mathbf{\varsigma \sigma} \overrightarrow{k} )_j
x_j} }}},
\end{equation}
where $(\mathbf{\varsigma \sigma} \overrightarrow{k} )_j$ is the
$j$th component wave number after the permutation operation
$\sigma$ and the sign-flipping operation $\varsigma$, ${(\mathbf{
\sigma} \overrightarrow{\kappa})}_1 \equiv \sqrt{k_0^2 -
{(\mathbf{ \sigma} \overrightarrow {k})_{1}}^2}$, which can be
regarded as an extra operator on top of the permutation operator
$\mathbf{\sigma}$, and $B( \mathbf{ \varsigma, \sigma})$ is the
superposition amplitude. Similarly, the wavefunctions in region
$R_3$ may be written as,
\begin{equation}\label{Bethe:Ansatz:wavefunction:3}
 \phi_{3}(\overrightarrow x)=
\sum_{\varsigma}{ \sum_{\mathbf{\sigma} \in \mathbf{G}}
{C(\mathbf{\varsigma, \sigma}) e ^{ i \sum_{j=1}^{N-1}{(\varsigma
\sigma \overrightarrow{k})_j x_j}} e
^{-(\mathbf{\sigma}\overrightarrow{ \kappa})_N x_N}}},
\end{equation}
where ${(\mathbf{\sigma} \overrightarrow{\kappa})}_N \equiv
\sqrt{k_0^2 - {(\mathbf{ \sigma} \overrightarrow{k})_{N}}^2}$ and
$C(\mathbf{\varsigma, \sigma})$ is the superposition amplitude.

From Eq.\eqref{define:kappa},  it is clear that if, for any i,
$k_i
> k_0$ then there will be a corresponding pure imaginary $\kappa_i$. From
Eq.\eqref {Bethe:Ansatz:wavefunction:2}, Eq.\eqref
{Bethe:Ansatz:wavefunction:3}, any pure imaginary $\kappa_i$ will
cause the Bethe ansatz wavefunction unnormalizable. Then from the
normalizability requirement stated at the end of Section
\ref{sec:intro}, we reason that \emph{a Bethe ansatz state is
bound if and only if all of the wave numbers are real and smaller
than $k_0$}. Since we are primarily interested in bound states,
from now on we implicitly mean bound state when we say Bethe
ansatz state, unless otherwise stated.

Once we get the Bethe ansatz wavefunctions written down, the rest
is straightforward. The main features of Eq. \eqref{eigen:problem}
are the singular $\delta$-function particle-particle interaction
and the nonzero potential step at the edge of the square well.
Bethe ansatz method elegantly treats both as boundary conditions.
The boundary conditions at $x_i -x_j = 0$ for $i,j=1,2,\cdots,N$
in regions of class $\mathcal A$ requires continuity of
wavefunctions on the one hand,
\begin{equation}
\label{bc:delta:continuity}\left.\psi \right|_{x_i = {x_j}^+} =
\left.\psi \right|_{x_i = {x_j}^-}
\end{equation} and certain discontinuity in their
first-derivatives on the other,
\begin{equation}
\left.\left[\frac{\partial \psi}{\partial x_i} - \frac{\partial
\psi}{\partial x_j}\right]\right|_{x_i = {x_j}^+} -
\left.\left[\frac{\partial \psi}{\partial x_i} - \frac{\partial
\psi}{\partial x_j}\right]\right|_{x_i = {x_j}^-} = 2 c
\left.\psi\right|_{x_i = x_j}. \label{bc:delta:derive:discon}
\end{equation}
The boundary conditions at $x_i = \pm x_0/2, i = 1,2,\cdots, N$
require continuity of both wavefunctions and their
first-derivatives:
\begin{eqnarray}
\label{bc:well-step:continuity}\left.\psi \right|_{x_i =
{x_0/2}^+}
&=& \left.\psi \right|_{x_i = {x_0/2}^-} \\
\label{bc:well-step:deriv:continuity}\left.\frac{\partial
\psi}{\partial x_i} \right|_{x_i = {x_0/2}^+} &=&
\left.\frac{\partial \psi}{\partial x_i} \right|_{x_i = {x_0/
2}^-}.
\end{eqnarray}
Plugging Eq.\eqref{Bethe:Ansatz:wavefunction:tau}, \eqref
{Bethe:Ansatz:wavefunction:2}, and \eqref
{Bethe:Ansatz:wavefunction:3} into Eq.\eqref
{bc:delta:continuity}, \eqref {bc:delta:derive:discon}, \eqref
{bc:well-step:continuity}, and \eqref
{bc:well-step:deriv:continuity} and including
Eq.\eqref{boson:exchange:symmetry}, we obtain the complete group
of equations for our original problem.

For the purpose of this article, it suffices to keep just the wave
numbers and eliminate all other unknowns. Doing so yields the
following secular equations for the $N$ wave numbers,
\begin{eqnarray}
\lefteqn{-x_0 k_j = - \pi I_j + 2 \sin^{-1} {\left(\frac{k_j}
{k_0}\right)} + \nonumber}\\ &&+\sum _{l=1, l \ne j}^N {\left[
\tan ^{-1} {\left({k_j + k_l}\right)} + \tan ^{-1} {\left({k_j -
k_l}\right)}\right]}, \label{eq:secular-equations}
\end{eqnarray}
where $j = 1,2,\cdots, N$, and $I=\{I_1, I_2,\cdots, I_N \}$ is a
set of preselected integers.

\begin{figure}[t]
\includegraphics[width=\figwidth] {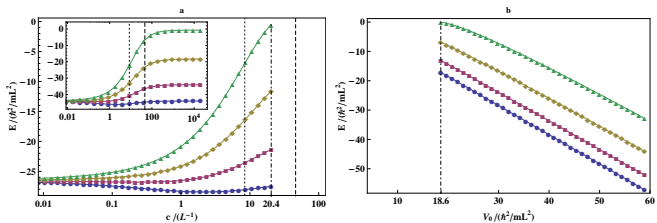}
\caption{(color online) Single particle energies of $4$ sodium
atoms in Bethe ansatz ground states. Trap size $L = 5\mu$m.
\textbf{a.} Dependence on interaction strengths ($c$). Trap depth
$V_0 = k_B \times 25$nK, where $k_B$ is Boltzmann's constant. The
dotdashed vertical line denotes the maximum interaction strength
above which no Bethe ansatz state of 4-boson system exists. The
other two vertical lines denote the interaction strengths of
sodium (dotted) and $^{87}$Rb (dashed) atoms at $\omega_\perp =
2\pi \times 150$ kHz and zero magnetic field. Inset, the trap
depth is lifted to $k_B \times 40$nK, at which condition all 4
atoms remain trapped to the Tonks limit. \textbf{b.} Dependence on
trap depths ($V_0$). Transverse trapping frequency $\omega_\perp =
2\pi \times 150$ kHz. Magnetic field is zero. The vertical line
(dotdashed) denotes the minimum trap depth below which no bound
state of 4-boson system exists. \label
{fig:BA-single-particle-energy}}
\end{figure}

Now a retrospect on the three regions we selected in the
coordinate space might make you wonder why we haven't included
more regions (and more equations), possibly with 2 or more
particles lying outside the trap area simultaneously. In reality,
this is possible, but only with small probability for deeply bound
state. The effect on the number-state condition should be
ignorable. Firstly, insofar as representative limiting cases
(strong interaction limit and deep trap limit), Bethe ansatz
solution agrees with known results (see Fig.
\ref{fig:BA-single-particle-energy}a). Secondly, the energy
spacings between single-particle levels are big near the strong
interaction regime where number state experiments take place most
likely and therefore the probability for more than one particle to
tunnel into the barrier is ignorably small at all times. For this
reason, we argue that the Bethe ansatz-based approach is a
sufficiently good approximation to our problem.

There are a few notable aspects of Eq.
\eqref{eq:secular-equations}. Firstly it is a transcendental
equation and so does not have analytic solution. Secondly, we must
pick a set of integers $I=\{I_1, I_2,\cdots, I_N \}$ before we
start the numerical computation. Apparently, not any set of
integers would lead to a physically meaningful solution. Then a
natural question to ask is what does. As argued in Appendix
\ref{append:validity:1}, we find that \emph{Eq.
\eqref{eq:secular-equations} yield valid solution if and only if
the set of integers $I$ are mutually distinct, somewhat similar to
the theorem in Ref. \cite{Yang:1967}.} A corollary of is that wave
numbers thus obtained have mutually distinct absolute values.
Because of this one-one correspondence, we use the set $I$ as the
\emph{quantum numbers} for the corresponding Bethe ansatz state.

\begin{figure}[t]
\includegraphics[width=\figwidth]{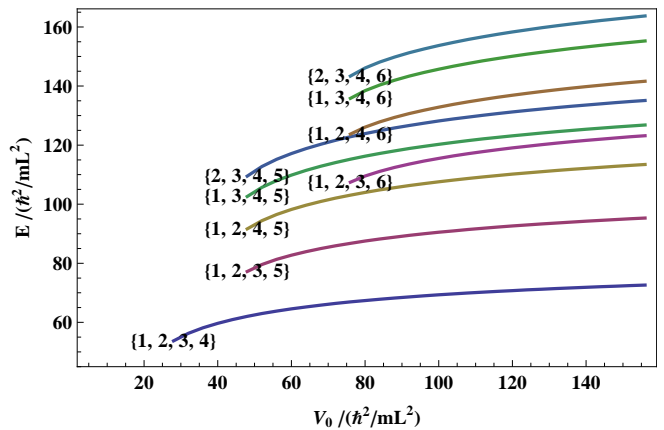}
\caption{(color online) Bethe ansatz states for sodium atoms. The
total energy of 4-particle bound states are plotted against trap
depth. The numbers at the beginning of each energy level are the
quantum numbers of the bound state. We shifted the energy zero to
the bottom of the trap for ease of presentation. Trap size $L =
5\mbox{ }\mu$m. Transverse trapping frequency $\omega_\perp =
2\pi\times 150$ kHz. \label{fig:excite-energies}}
\end{figure}

Besides knowing what set of integers are valid quantum numbers, we
need further identify the ground state and the first-excited
state. As argued in Appendix \ref{append:energy:level:order:1}, we
find that the ground state of an $N$-boson system has the quantum
number $\{1,2, \cdots, N\}$; the first-excited state has the
quantum number $\{1, 2, \cdots,N-1, N+1\}$ (see Fig.
\ref{fig:excite-energies}).

Without loss of generality, we re-organize the set of wave numbers
such that $0 < k_1< k_2< \cdots< k_N< k_0$. We define \emph{the
$j$th single particle energy} as $e_j=-{\kappa_j}^2 /2$, for $j =
1,2,\cdots, N$. Fig. \ref{fig:excite-energies} gives total
energies of the low-lying Bethe ansatz states.

\section{Number states \label{sec:number-state}}
\begin{figure}[t]
\includegraphics[width=\figwidth]{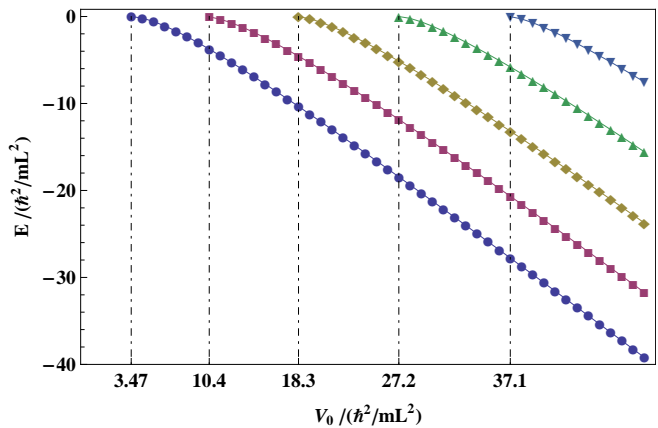}
\caption{(color online) Ionization thresholds of sodium atoms with
all parameters fixed except trap depth (we use $\hbar^2/mL^2$ as
energy unit for convenience). Only highest single particle
energies of the Bethe ansatz $N$-boson states are shown with $N =$
2 (circle), 3 (square), 4 (diamond), 5 (upright triangle), and 6
(invert triangle).  Trap size $L= 5\mbox{ }\mu$m; Transverse
trapping frequency $\omega_\perp = 2\pi \times 150$ kHz. The
ionization thresholds (with the current numeric calculation step
size) are marked with vertical lines (dot-dashed).
\label{fig:lower-bound-N-boson-state}}
\end{figure}

We now apply the results of the previous sections to the problem
of number state production. We can safely assume no atom with
positive energy presents in space near the trapped area. In
reality, if an atom acquires positive energy, it would be quickly
swept out of the vacuum chamber. Therefore, it is safe to assume
that the states in the continuum spectrum are virtually unoccupied
all the time.

As dictated by Bethe ansatz, for some given trap parameters (depth
$V_0$, trap size $L$, scattering length $a$, and transverse
trapping frequency $\omega_\perp$), $N$ bosons can be contained in
the trap if and only if there is an $N$-boson Bethe ansatz state.
The energy levels for an $N$-boson system has been computed by
Eq.\eqref{eq:secular-equations} numerically. As an example, Fig.
\ref{fig:BA-single-particle-energy} shows a $4$-boson Bethe ansatz
state that ceases to exist in certain regions of the parameter
space. The Bethe ansatz state can only exist \emph{down} to a
certain trap depth in panel \textbf{a} and can only exist
\emph{up} to certain interaction strength in panel \textbf{b},
provided that all other parameters are held unchanged. We call the
maximum number of particles that can be contained in the trap the
\emph{trap capacity}. The trap capacity puts an upper bound on the
number states for a given point in the parameter space. The whole
parameter space is thus partitioned into zones of certain trap
capacities. We define the boundaries of these partitions as the
\emph{ionization threshold}, because as we cross the boundaries
from higher trap capacity side to lower capacity side
adiabatically, the system state changes from stable to unstable
and must release some particle(s). In Fig.
\ref{fig:lower-bound-N-boson-state} we show the ionization
threshold for a system of 2, 3, 4, 5 and 6 bosons.

\begin{figure}
\includegraphics[width=\figwidth]{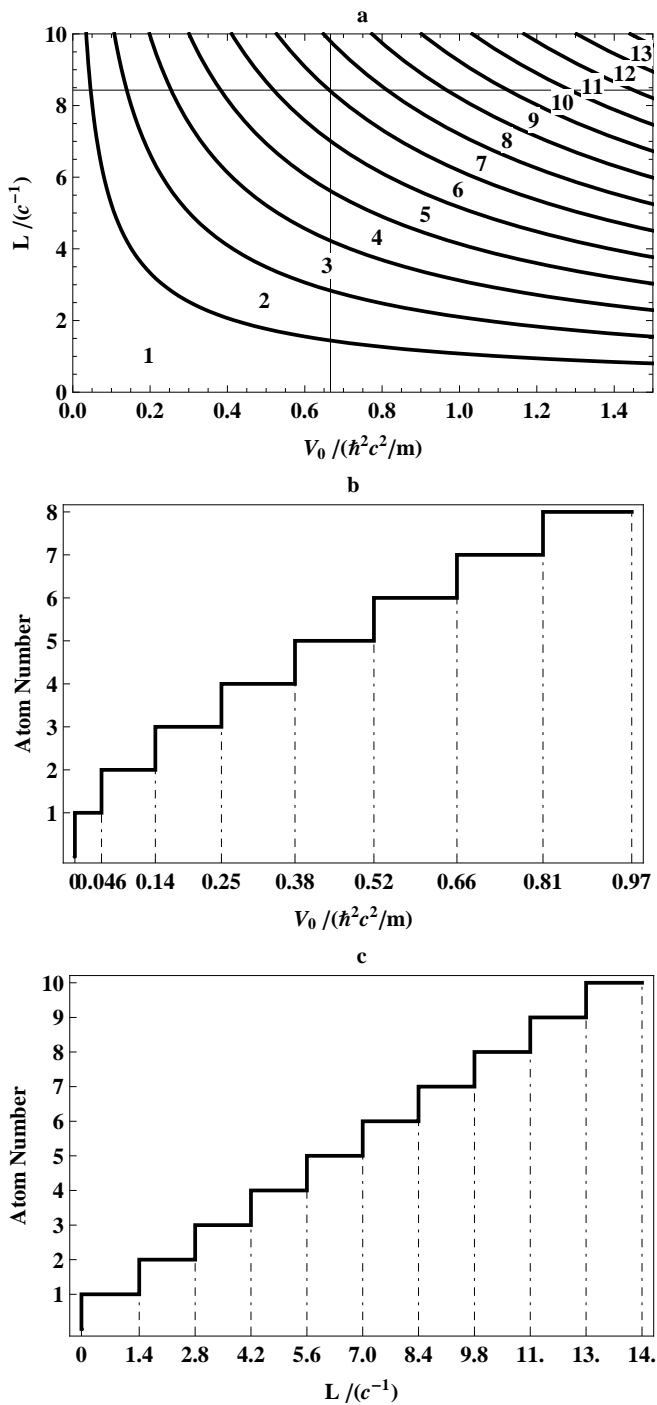}
\caption{Map of number states and the ionization thresholds for
sodium atoms in 1D optical trap in adiabatic limit. Transverse
trapping frequency $\omega_\perp = 2\pi\times 150$ kHz and zero
magnetic field are assumed. Note that $c^{-1}$ and $\hbar^2 c^2/m$
are used to make the axes dimensionless. \textbf{a.} Contour plot
of number states as function of trap depth and size; \textbf{b.}
and \textbf{c.} cross-sectional cut views at the indicated trap
size ($5\mu$m) and depth ($4$nK) (solid lines) of a. The ticks on
horizontal axes give the calculated ionization thresholds.
\label{fig:number-vs-pot-wid}}
\end{figure}

Now that we have a clear upper limit, the trap capacity, on the
number state that may be present for a given trap and other
physical parameters. It remains questionable whether or not the
trap capacity can be reached. The adiabatic laser culling
technique developed in Ref. \cite{Dudarev:2007} and the
simulations in Tonks-Girardeau region made in Ref.
\cite{ACampo:2008-2009} seem to suggest that it is possible to
reach the trap capacity with the ultra-cold technique developed in
Ref. \cite{Chuu:2005,Meyrath:2005}.

The starting point is a almost-pure Bose Einstein condensate (BEC)
that is optically trapped. Ignore excitation effects for now, it
is useful to view the process from the angle of quantum optics and
regard the state of the BEC as a coherent state
\cite{Andrews:1997}. A coherent state is essentially a
superposition of Fock states with a Poisson-distribution in the
boson numbers. As we adiabatically change experimental parameters
from partitions with higher trap capacity to lower capacity
targeting some number state, the Bethe ansatz solution puts a
tighter and tighter restriction on the maximum number state. The
system state thus undergoes two changes side by side: 1, more and
more high-energy atoms are forced out; 2, more and more
high-number Fock states are quantum mechanically `projected' out
of the system state (resulting in the so-called \emph{squeezed
state}). Each of these two changes has its distinctive effect on
the system state: the first leads to smaller and smaller average
particle number $\overline{N} = \left\langle N\right\rangle$
whereas the second leads to a reduction in the number uncertainty
$\sigma^2 = \left\langle N^2 - \overline{N}^2\right\rangle$. Under
optimal experimental conditions, the process continues until at
some point, while the average number $\left\langle N\right\rangle>
0$, the number uncertainty $\sigma \approx 0$. A rigorous
simulation of this would require calculating the value
$\frac{\sigma}{\overline{N}}$ as a function of time in a dynamic
process and is certainly beyond the scope of this article. In Fig.
\ref{fig:number-vs-pot-wid}, we show trap capacities and
ionization thresholds as functions of trap depth and size. The
interaction strength is implicit in the unit we adopted.

There are several ways to tune the physical parameters to achieve
the above goals. In previous references
\cite{Dudarev:2007,ACampo:2008-2009}, only culling (changing trap
depth), squeezing (changing trap size), or some combinations of
the two are discussed. In certain circumstances, we propose that
atom-atom interaction strength $c$ be possibly tuned to supplement
the production of number states. In view of the intrinsic
limitations in tuning the trap parameters, it's possible that
tuning of interaction strengths could play a key role in number
state experiments.

The path to a number state becomes clear now. By tuning the
physical parameters of the 1D optical trap adiabatically, we force
the ultracold atom sample through a series of quantum collapses
until it eventually reaches the desired Fock state with some
acceptable fidelity. Ideally, the course connecting the starting
point and a targeted Fock state consists of a series of states
(the Bethe ansatz states) with well-defined particle number. But
in reality, there are always some \emph{elementary excitations},
which is defined as any deviation from the ideal adiabatic course.
Possible elementary excitations include occupations of excited
Bethe ansatz state (of the same particle number), earlier
ionizations (loss of particles before reaching the Bethe ansatz
ionization threshold), and simultaneous ionizations of more than
one particles.

\begin{figure}[b]
\includegraphics[width=\figwidth] {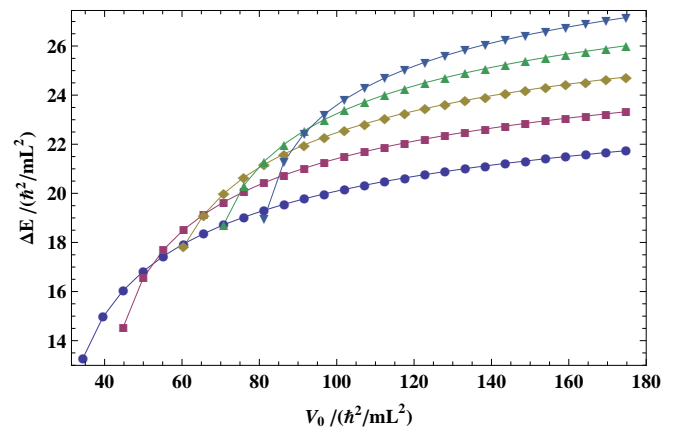}
\caption{(color online) Excitation energy gaps between ground and
first-excited states as function of trap depth for 2 (circle), 3
(square), 4 (diamond), 5 (upright triangle), and 6 (invert
triangle) sodium atoms. Trap size is $5\mbox{ }\mu$m. \label
{fig:excite-energy-gap}}
\end{figure}

We now analyze the effects of excitations. Abrupt changes in the
trapping potential tend to introduce extra terms into the system
density matrix. As the system gets near an ionization threshold,
the system becomes particularly delicate, since the particle with
the highest energy can tunnel further away from the center of the
trap and thus external disturbance has bigger exciting effect on
the system. Moreover, immediately after the ionization threshold
is passed, the system density matrix is subject to various
excitations due to wavefunction collapses. These excitations are
crucial to the fidelity of Fock state production, since they cause
significant reversion in the number uncertainty of the final
state. A characteristic measurement of tendency of excitation is
the \emph{energy gap}, $\Delta$, which is define as the difference
between total energies of ground and first-excited Bethe ansatz
states (if both exist). According to our calculation, they are on
the order of a few $k_B\times10$nK (see Fig.
\ref{fig:excite-energy-gap}). $\Delta$ puts restrictions in
two-fold. Firstly, the temperature must be maintained lower than a
few $10$nK, otherwise, fidelity could be endangered due to thermal
excitation. Furthermore, the energy gap puts a requirement on the
adiabaticity condition \cite{adiab:theorem}: the culling speed
must be much smaller than $\frac{\Delta^2}{\hbar}$ to maintain a
relatively high fidelity. To give a sense of number, we consider
the culling of trapping potential from the ionization threshold of
3 particles down to that of 2 particles at trap size of $5\mu$m
and transverse trapping frequency $\omega_\perp = 2\pi
\times150$kHz. According to our calculation, the minimum time
required to complete this portion of the culling should be no less
than $0.3$ms to be considered as adiabatic.

\section{conclusion}
In conclusion, we calculated the conditions for number states of
ultracold atoms in 1D optical trap with the Bethe ansatz approach.
We charted ionization thresholds in the parameter space. We also
discussed the quantum mechanical processes in producing number
states in the ideal case and the effect of excitations.

\begin{acknowledgments}
We acknowledge support from the NSF and from the R.A. Welch
Foundation. M.G.R. acknowledges support from Sid W. Richardson
Foundation. We thank Greg Fiete for inspiring suggestions. S.P.W
would like to thank Tongcang Li, Hrishikesh Kelkar, Shengyuan Yang
for the useful discussions.
\end{acknowledgments}

\appendix

\section{Valid Bethe ansatz solutions \label{append:validity:1}}
First of all, we need a change of unit to make the dependence of
secular equation on interaction strength explicit. (See Section
\ref{sec:intro} for the previous choice of unit.) To that end, we
choose the trap length $L$ as the length unit, $\hbar^2/mL^2$ as
the energy unit. Then the secular equation
Eq.\eqref{eq:secular-equations} is transformed to
\begin{eqnarray}\label{eq:transf:secular-equations}
\lefteqn{-k_j L = \pi I_j + 2 \sin^{-1} {\left(\frac{k_j}
{V_0}\right)} + } \nonumber \\ &&\sum _{l \neq j} {\left[ \tan
^{-1} {\left(\frac{k_j + k_l}{c}\right)} + \tan ^{-1}
{\left(\frac{k_j - k_l}{c}\right)}\right]},
\end{eqnarray}

Now Eq.\eqref{eq:transf:secular-equations} depends both on the set
of integers $I$ and the interaction strength $c$. For the mere
purpose of solving Eq.\ref{eq:secular-equations}, any set of
integers $I$ may be used. Note that the choice of integers is
discrete and is usually enumerable while that of the interaction
strength $c$ is continuous and non-enumerable. Based on this fact,
we claim that if a given set of integers lead to valid solution at
\emph{some} interaction strength, it does so at \emph{any}
interaction strength. Particularly, the solution in the weak
interaction region should approach that of the non-interacting
case as $c \to 0$. Solution of the latter is a well-taught
exercise in many quantum mechanics textbooks.

Thus, a `promising solution' to
Eq.\eqref{eq:transf:secular-equations} should converge onto that
of the $c=0$ case and we can use the known solutions to reject
spurious `solutions' for the interacting cases. With a few
examples of $N$ and upper limits $M$, we exhaust all the
combinadics of $N$ numbers from the range $[0,M]$ and solve
Eq.\eqref{eq:transf:secular-equations} with each combination. Our
experiments show that in the limit $c \to 0$, a solution
approaches that of $c = 0$ if and only if the set $I$ consists of
positive and mutually distinct integers. Furthermore, we also find
that the wave numbers in the solution are mutually distinct if and
only if the integers in the set $I$ are mutually distinct.
\section{Order of Bethe ansatz states \label{append:energy:level:order:1}}
With the valid solutions found in Appendix
\ref{append:validity:1}, it still left to determine which one is
the ground state and which is the first excited state, and so on.
In a similar manner as in Appendix \ref{append:validity:1}, we
will argue based on intuition that for any given $N$, the set $I =
\{1,2,\cdots, N\}$ leads to the ground state and the set $I
=\{1,2,\cdots, N-1, N+1\}$ to the first excited state.

Our clue comes from the strong interaction limit. As is well
known, in the strongly interacting limit the particles behave as
fermions \cite{Girardeau:1960}. Thus the ground state of our
system must be like that of a degenerate fermion system. Our
numerical calculations show that solving
Eq.\eqref{eq:transf:secular-equations} with the set $I =
\{1,2,\cdots, N\}$ leads to a solution with energy that approaches
that of the ground state of the degenerate fermion system in the
limit $c\to \infty$. Furthermore, the solution obtained with the
set $I =\{1,2,\cdots, N-1, N+1\}$ approaches that of the
first-excited state of the same system in the same limit. We thus
established what are ground state and first excited state for our
system in general. However, there is little we could say beyond
that. Within limit of calculation error, our experiment is not
conclusive about which is the second excited states. It might be
$1,2,\cdots, N-1, N+2$, $1,2,\cdots, N-2, N, N+1$, or still other,
depending on $N$ and the trap parameters. In general the order in
the energy level of a Bethe ansatz state depends on both the
highest quantum number and the total of these quantum numbers. For
complete ordering, one need something as what Hund's Rule is in
atomic physics. For our paper, it suffices to know just the ground
state and first excited state.


\begin{thebibliography}{widest-label}
\bibitem{comment1} There are other technical reasons in tuning
the interaction strength. To facilitate efficient loading of Bose
Einstein condensate into the optical box, the transverse trapping
frequency is often relaxed in the beginning and tightened up
toward the end to promote atom-atom interaction. This process
often gets involved with that of the number-state production.
Therefore, we need to study the connections across various
interaction regimes.
\end{thebibliography}

\end{document}